# Charged domain boundaries stabilized by translational symmetry breaking in the hybrid improper ferroelectric $Ca_{3-x}Sr_xTi_2O_7$

Hiroshi Nakajima[1, *], Kosuke Kurushima[2], Shinya Mine[3], Hirofumi Tsukasaki[1], Masaya Matsuoka[4], Bin Gao[5], Sang-Wook Cheong[6], and Shigeo Mori[1]

[1]*Department of Materials Science, Osaka Prefecture University, Sakai, Osaka 599-8531, Japan.*
[2]*Toray Research Center, Ohtsu, Shiga 520-8567, Japan.*
[3]*Institute for Catalysis, Hokkaido University, N-21, W-10, Sapporo 001-0021, Japan.*
[4]*Department of Applied Chemistry, Osaka Prefecture University, Sakai, Osaka, 599-8531, Japan.*
[5]*Department of Physics and Astronomy, Rice University, Houston, Texas 77005-1827, USA.*
[6]*Rutgers Center of Emergent Materials, Rutgers University, Piscataway, New Jersey 08854, USA.*
*e-mail: nakajima@mtr.osakafu-u.ac.jp*

## Abstract

**Charged domain walls and boundaries in ferroelectric materials display distinct phenomena, such as an increased conductivity due to the accumulation of bound charges. Here, we report the electron microscopy observations of atomic-scale arrangements at charged domain boundaries in the hybrid improper ferroelectric $Ca_{2.46}Sr_{0.54}Ti_2O_7$. Like in the prototype improper ferroelectric $YMnO_3$, we find that charged domain boundaries in $Ca_{2.46}Sr_{0.54}Ti_2O_7$ correspond to out-of-phase boundaries, which separate adjacent domains with a fractional translational shift of the unit cell. In addition, our results show that strontium ions are located at charged domain boundaries. The out-of-phase boundary structure may decrease the polarization charge at the boundary because of the ferrielectric nature of $Ca_{2.46}Sr_{0.54}Ti_2O_7$, thereby promoting the stabilization of the charged state. By combining atomic-resolution imaging and density-functional theory calculations, this study proposes an unexplored stabilization mechanism of charged domain boundaries and structural defects accompanying out-of-phase translational shifts.**

# Introduction

The physical properties of ferroelectric domain walls and boundaries are sometimes different from those of domains themselves[1–3]; examples include ferroelectricity at the antiphase boundary of an antiferroelectric[4] and polar ferroelectric structures at ferroelastic boundaries[5]. Domain walls can have higher or lower conductivity than domains, a feature that could be used in high-density storage and switching devices[2,6,7]. In particular, charged domain walls, in which electric polarization vectors are opposed between adjacent domains, have attracted considerable attention because these walls could induce an enhanced conductivity in ferroelectric materials[8]. Charged domain walls are energetically unfavorable[9]; however, they are obtained by combining other structural parameters because ferroelectricity is not a primary order parameter in materials known as improper ferroelectrics[10]. For instance, improper ferroelectric hexagonal manganese oxides ($RMnO_3$: $R$ = Sc, Y, or Ho–Lu) exhibit peculiar cloverleaf domains with head-to-head or tail-to-tail domain walls involving polarization-dependent conductivity[11–13], which serves as a bias switching for devices[14].

Recently, perovskite oxides whose ferroelectricity is induced by more than two nonpolar octahedral tilts (hybrid improper ferroelectricity) have been extensively studied because of the possibility of achieving giant magnetoelectric effects[15]. Through the exploration of hybrid improper ferroelectrics, numerous peculiar charged domain walls were found in $Ca_{3-x}Sr_xTi_2O_7$ ($x$ = 0.54)[16]. This material has a layered perovskite structure with the $A2_1am$ (No. 36) space group, which comprises a stack of perovskite and rock-salt layers, as shown in Figure 1. The structure is classified as a Ruddlesden–Popper phase ($Ca_{n+1}Ti_nO_{3n+1}$) with $n$ = 2. The existence of two nonpolar $TiO_6$ tilts leads to Ca/Sr displacements along the [100] axis, inducing ferroelectricity accompanying charged domain walls. Notably, the number of charged domain walls in the crystal is substantially increased through Sr substitution[16]. However, the microscopic structure of the charged domain walls remains elusive. In particular, the effect of Sr substitution has not been revealed, which motivated us to perform atomic-resolution scanning transmission electron microscopy (STEM) with energy-dispersive X-ray spectroscopy (EDS).

Another motivation for this work is to reveal the role of crystallographic defects in $Ca_{3-x}Sr_xTi_2O_7$. For $RMnO_3$, recent extensive studies have revealed that antiphase boundaries, which separate adjacent domains with a translational shift, have played an important role in forming charged domain walls[17–20]. Translation domains (i.e., domains in which the translational symmetry is broken) are formed due to the trimerization of $MnO_5$, which results in the stabilization of charged ferroelectric domain walls[11].

However, in the improper ferroelectric $Ca_{3-x}Sr_xTi_2O_7$, the existence and effect of translation domains have not been observed at the atomic scale.

Here, we observed atomic-scale local structures to unveil the stabilization mechanism of the numerous charged defects in $Ca_{2.46}Sr_{0.54}Ti_2O_7$. The experimental results show that some charged boundaries are out-of-phase boundaries (translational boundaries) that contain Sr ions at the boundaries. Charged out-of-phase boundaries separate both the ferroelectric polarization directions and the crystallographic domains with phase shifts. Notably, an out-of-phase boundary is defined as a boundary between two regions of a crystal displaced by a fractional translation of a unit cell[21–23]. In particular, an out-of-phase boundary is called an antiphase boundary when the displacement phase is π (i.e., half of the lattice constant). Also note that, in this paper, a domain boundary represents a crystallographic defect that separates two ferroelectric domains with opposite polarization directions while a domain wall means a ferroelectric domain boundary without a compositional change. The out-of-phase displacement allows local electric polarization vectors of adjacent domains to align in the same direction at the boundaries, which stabilizes the charged domain boundaries. This study proves that charged domain boundaries are correlated with crystallographic defects in $Ca_{2.46}Sr_{0.54}Ti_2O_7$.

**Results and discussion**

**Dark-field images of ferroelectric domains.** To explore charged structural defects in the hybrid improper ferroelectric $Ca_{2.46}Sr_{0.54}Ti_2O_7$, its $(1\bar{1}0)$ plane was observed. Observations from the $[1\bar{1}0]$ direction are expected to be crucial because numerous charged defects run along this direction, which is at 45° to the macroscopic polarization directions[16]. Figure 2a,b shows dark-field images in the $(1\bar{1}0)$ plane, which depict the characteristic defects and the ferroelectric domains with polarization pointing toward the left- and right-hand side of the image, as denoted by the blue and red arrows, respectively. The image was obtained under a two-beam condition by exciting a Bragg reflection, which causes the breakdown of Friedel's law[24–26]. The imaging method identifies ferroelectric domains as bright and dark areas when the ferroelectric polarization is parallel and antiparallel to the Bragg reflection used in the dark-field imaging, respectively. Thus, the images demonstrate that the specimen has numerous neutral 180° and charged head-to-head and tail-to-tail ferroelectric domains. Curved charged structural defects exist, as reported in Ref[27]. However, sharp straight charged domain boundaries are frequently formed, as indicated by the squares labeled 3a and 3b. Moreover, broad domain boundaries are also visualized (labeled S6a). Atomic-resolution observations were performed to reveal these characteristic defects.

**Atomic-resolution observations in charged domain boundaries.** Figure 3a shows the high-angle annular dark-field STEM (HAADF-STEM) image of the corresponding area indicated in Figure 2. The lattice structure can be described using five ions between the rock-salt layers, as depicted in Figure 1: The Ca, Ti, Sr, Ti, and Ca ions are arranged along the [001] axis. The dark lines correspond to the rock-salt layer, since the rock-salt layer has less atomic density than the perovskite layer. Noticeably, the left region is misaligned compared with the right region. The boundary is pinched at an intergrowth with seven layers (yellow arrowhead, Figure 3b), which agrees with the contrast of the region labeled 3b in Figure 2. A two-layer displacement occurs at the boundary, as indicated by the dotted line in Figure 3c. The displacement of the layers corresponds to $0.1889c$ along the [001] direction (where $c$ is the lattice constant for the $c$ axis), which demonstrates that the boundary is out of phase. Moreover, the boundary is sharp at the atomic scale. The EDS maps also reveal this boundary structure. In the matrix domains, the EDS maps show that Ca ions occupy the alkaline-earth-metal sites of the rock-salt layer (Figure 3d), Sr ions occupy the alkaline-earth-metal sites of the perovskite layer (Figure 3e), and Ti ions occupy the metal sites of the oxygen octahedron (Figure 3f). The atomic arrangement shown in Figure 3g agrees with the crystal structure illustrated in Figure 1. However, at the out-of-phase boundary, the EDS maps of Figure 3d–g show that the alkaline-earth-metal site of the perovskite layer coincides with that of the rock-salt site. As indicated by the dotted lines, the Sr layers (red dots) are located adjacent to the Ca layers (green dots) at the boundary. In the dark-field image of Figure 2, the domain contrast indicates a tail-to-tail charged boundary, delineating a sharp straight line: The polarization is inverted across the boundary, as indicated by the $P_s$ arrows, and the boundary lies at an angle of approximately 80° to the polarization in the two domains. These results demonstrate that the charged domain boundary coincides with the out-of-phase boundary. The charged boundary that resembles a domain wall in the dark-field image of Figure 2 is technically a charged defect. Furthermore, the HAADF-STEM image shows brighter dots at the boundary, indicating the presence of heavier ions. The EDS maps show that Sr ions are segregated at the boundary. From these observations, it can be inferred that the formation of charged domain boundaries is influenced by the out-of-phase boundaries that contain Sr ions. The increase in the number of charged domain walls through Sr substitution[16] can be explained by the presence of Sr ions at the charged boundaries because Sr ions are likely to form charged boundaries, as observed in this study. From the microscopy study, it was also found that another out-of-phase boundary with one-layer shift coincides with a head-to-head charged boundary involving Sr segregation (Supplementary Figure 1), demonstrating that out-of-phase boundaries are important in the formation of charged domain boundaries.

It is important to explain why charged domain boundaries are formed at out-of-phase boundaries. Ferroelectric domains are formed as a consequence of the interplay between the wall energy and the depolarization fields induced by the bound charges due to the electric polarization[28]. Charged domain walls (boundaries) are energetically unfavorable compared with neutral domain walls because they create bound charges, which increase the electrostatic energy of the wall. Bulk $Ca_{2.46}Sr_{0.54}Ti_2O_7$ is a ferrielectric material whose polarization is caused by the combination of the two left displacements and one right displacement in perovskite and rock-salt layers, respectively (Figure 1). When the structure is displaced by $0.1889c$ at the out-of-phase boundary, the oppositely polarized layers show the same displacement directions, as indicated by the dashed lines in Supplementary Figure 2. The bound charge $\rho$ is expressed as $\rho = -\nabla \cdot P_s$, where $P_s$ is the electric polarization. Since the bound charge is caused by the divergence of the polarization, parallel polarization directions reduce the local bound charge. Therefore, the configuration observed is energetically favorable, which possibly explains the origin of the charged out-of-phase boundary. Moreover, first-principles calculations show that the polarization vectors differ from layer to layer[27,29]. The ferrielectricity of $Ca_{2.46}Sr_{0.54}Ti_2O_7$ can be described by a stack of parallel- and antiparallel-layered polarizations as shown in Supplementary Figure 3. Each layered polarization vector points toward opposite directions when the charged domain wall is formed. However, some layers have the same polarization directions if the domain is shifted by two layers ($0.1889c$) or one layer ($0.122c$), which are experimentally observed in Figure 3 and Supplementary Figure 1, respectively. Due to the bound charge, these configurations should be more stable than boundaries without a translational shift. This mechanism is possible because of the ferrielectric nature of $Ca_{2.46}Sr_{0.54}Ti_2O_7$, which exhibits layered polarization vectors with opposite directions. Hence, this phenomenon should be characteristic of layered perovskite structures showing ferrielectricity.

Furthermore, the difference between $Ca_{2.46}Sr_{0.54}Ti_2O_7$ and $YMnO_3$ can also be explained. In $YMnO_3$, the trimerization of $MnO_5$ causes cloverleaf charged ferroelectric domain walls due to the structural phase transition[11]. The cloverleaf domains comprise ferroelectric domain walls that coincide with antiphase boundaries because ferroelectric walls with antiphase boundaries have lower energy than only ferroelectric walls or antiphase boundaries[11,17]. Although $YMnO_3$ and $Ca_{2.46}Sr_{0.54}Ti_2O_7$ share several properties, such as the formation of charged defects, translation domains, and improper ferroelectricity due to structural order parameters, the origin of the charged domain boundaries accompanying the translation domains in $Ca_{2.46}Sr_{0.54}Ti_2O_7$ differs from that in $YMnO_3$. The out-of-phase boundaries in $Ca_{2.46}Sr_{0.54}Ti_2O_7$ are not induced by the phase transition from the *I4/mmm* to the *A2$_1$am* space group because the shear structure of the layer observed in Figure 3 is not related to the symmetry of the phase transition. This is significantly different from the antiphase boundary due to the transition in YMnO$_3$. The result obtained for Ca$_{2.46}$Sr$_{0.54}$Ti$_2$O$_7$ is also supported by the presence of Sr ions at the boundaries because the phase transition will not move Sr, which would take their positions after the material has become crystalline. Thus, the out-of-phase boundaries should exist in the *I*4/*mmm* paraelectric phase. Therefore, as the temperature is decreased, charged boundaries are selectively formed at the out-of-phase boundary positions because of the above-explained energy gain at the ferroelectric transition.

Notably, in planar defects with dislocations, one domain is shifted with respect to the adjacent domain, and this shift is characterized by a displacement vector[30,31]. If the displacement vector is a translation vector of the disordered structure, the boundary is known as an antiphase boundary. By contrast, if the displacement vector is not a lattice translation vector, the boundary is known as a stacking fault. As shown in Figure 3, the domains are shifted by fractions of the unit cell, demonstrating that the displacement vector is not a lattice translation vector. Thus, the out-of-phase boundary observed in Figure 3 is also defined as a stacking fault. This study thus reveals charged stacking faults in ferroelectrics, which have not been reported in layered perovskite structures. The displacement vector of the stacking fault could not be determined here because of the diffraction contrast due to the ferroelectricity and the low symmetry of the structure via dark-field imaging. However, Figure 3 shows that the displacement in the ($1\bar{1}0$) plane occurs only along the [001] direction. Besides, the HAADF-STEM images shown in Supplementary Figure 4 indicate that no displacement can be seen in boundaries for which the dark-field image and the STEM images display boundary contrast when the (001) plane is investigated. These images also show that this type of boundary runs along the [$1\bar{1}0$] direction, and the boundary is a planar defect.

**Charged domain boundaries with rotated out-of-phase domains.** Another structural defect is observed in Figure 4a, in which a head-to-head boundary is formed. Figure 4b–f shows magnified HAADF-STEM images and corresponding EDS maps around the defect. Similar to Figure 3, the rock-salt layers are misaligned by two layers, indicating the formation of an out-of-phase boundary. However, the boundary comprises four ions in width. Besides, the presence of Sr and Ti ions indicates that these regions form a perovskite layer (SrTiO$_3$) because Sr ions preferentially occupy the sites of the perovskite layer, as indicated by the density-functional theory calculations (Supplementary Note 5), and the rock-salt layer is not formed. At the edges of the boundary, dark lines can be observed parallel to the [001] axis in the HAADF-STEM image, whose contrast is characteristic of the rock-salt layer. The elemental maps in Figure 4c demonstrate that these edges comprise Ca ions parallel to the [001] axis. Since this region has Ca layers that correspond to the rock-salt layer along the [001] direction, the region is considered a 90°-rotated domain whose [001] axis is perpendicular to the crystal [001] axis. The dark-field image in Figure 4 indicates that this boundary has a head-to-head charged configuration, as illustrated by the arrows in Figure 4a. These $SrTiO_3$ regions seemingly stabilize the charged domain boundary. Contrary to $Ca_3Ti_2O_7$, the perovskite structure of simple cubic $SrTiO_3$ is nonpolar. Thus, it is energetically more favorable to form the $SrTiO_3$ perovskite structure at the charged domain boundary. The presence of a nonpolar region increases the stability of the charged domain boundary because these regions separate oppositely polarized domains. Thus, these results reveal that the local structures and origins of the charged domain boundaries in this case are different from those illustrated in Figure 3, although they are formed in the same crystal.

Finally, the effects of Sr substitution in $Ca_{2.46}Sr_{0.54}Ti_2O_7$ are explained. Previous polarization–hysteresis measurements have revealed that the spontaneous polarization magnitude decreases as the Sr content increases[16]. Such a reduction occurs because the structure of Sr-substituted $Ca_{3-x}Sr_xTi_2O_7$ is nonpolar (space group $P4_2/mnm$ for $0.915 < x < 1$ and $I4/mmm$ for $x > 1$)[32]. Accordingly, the increased Sr content reduces the octahedral tilt in the crystal structure, which decreases the polarization. Additionally, Sr substitution results in a decrease in the polarization through the formation of $SrTiO_3$ regions in a crystal. Our observations show out-of-phase boundaries that comprise the $SrTiO_3$ perovskite structure. Furthermore, a large perovskite region with higher Sr content than the matrix region was observed as shown in Figure 2a and Supplementary Figure 6. The areas with the $SrTiO_3$ perovskite structure have no polarization because of the nonpolar simple cubic structure of $SrTiO_3$. Hence, the extended perovskite region reduces the spontaneous polarization, which is defined as the electric polarization per volume.

**Conclusions**

The dark-field and HAADF-STEM images prove that the crystal has distinct structural boundaries and defects. The elemental maps show the presence of out-of-phase boundaries, which contain Sr ions. Furthermore, it was demonstrated that charged domain boundaries correspond to out-of-phase boundaries because the ferrielectric structure compensates for the polarization charge at the boundaries by means of translational shifts. Notably, this study reveals that some charged boundaries considered as charged domain walls in a previous work[16] could actually be charged defects. The structural differences between charged walls and defects are evidenced in their macroscopic behaviors. For example, charged domains walls can be moved via an external electric field, whereas domains separated by charged defects remain stationary under an electric field higher than the coercive field because they are pinned by Sr atoms. Besides, in charged defects of ferroelectrics, the valence states of cations are found to be different from those of domains[33]. Thus, the charged out-of-phase boundaries presented in this study may have different valence states of Sr and Ti ions at the boundaries. These properties will be explored in future works. Our results provide mechanistic insights into the microscopic structure of boundaries and the Sr substitution effect on the properties of the layered perovskite oxide $Ca_{3-x}Sr_xTi_2O_7$. Moreover, this study shows that atomic-resolution elemental mapping is important for understanding structural defects and macroscopic properties in ferroelectric materials.

**Methods**

**Scanning transmission electron microscopy**

Atomic-resolution HAADF-STEM and EDS-STEM mapping were performed using a transmission electron microscope (JEM-ARM200F, JEOL Co., Ltd., Japan) equipped with a spherical aberration corrector and double silicon drift detectors. The acceleration voltage, probe semiangle, and current were 200 kV, 22 mrad, and 60 pA, respectively. The solid angle for the whole collection system was ~1.96 Sr. in the EDS mapping. The angular detection range of the HAADF detector for scattered electrons was 90–170 mrad. EDS-STEM images were processed using Fourier filtering for noise reduction[34]. The single crystal was grown via the floating zone method. Thin specimens along the $(1\bar{1}0)$ plane were prepared via focused ion beam at 30 kV up to a thickness <100 nm. Subsequent Ar-ion milling at 4 kV and an incident angle of 8° was used to remove the damage from the specimen and reduce the specimen thickness.

The polarization direction vector $\boldsymbol{P_s}$ of each ferroelectric domain was determined from the contrast of the dark-field images, which were captured under two-beam conditions[24,25]. In these conditions, the intensity of a Bragg reflection $\boldsymbol{g_{hkl}}$ is different from that of a reflection with space inversion $\boldsymbol{g_{\bar{h}\bar{k}\bar{l}}}$, and the breakdown of Friedel's law occurs. Consequently, the contrast in the ferroelectric domain can be expressed as $\boldsymbol{P_s} \cdot \boldsymbol{g_{hkl}}$: The dark-field image indicates ferroelectric domains as bright areas when the Bragg reflection is parallel to the polarization. Conversely, ferroelectric domains are depicted as dark areas when the reflection is antiparallel to the polarization. The contrast disappears when the polarization is perpendicular to the reflection. The disappearance of the contrast was confirmed by using the 008 reflection, as shown in Supplementary Figure 7.

**Density-functional theory calculations**

All calculations were conducted based on the spin-polarized density-functional theory using the *Vienna ab initio* simulation package (VASP 6.1.0)[35]. The core–valence electron interaction was described using the projector augmented wave method[36], and the exchange-correlation part was treated within the Perdew–Burke–Ernzerhof generalized gradient approximation[37]. The on-site Coulomb interaction ($U_{eff}$) of the Ti ions was set to 3 eV[38]. The cutoff energy for the plane-wave basis was 550 eV. The atomic positions shown in Figure 1 were relaxed to obtain the most stable structure: The Hellman–Feynman forces on every atom converged to be <0.01 eV Å$^{-1}$. Besides, convergence concerning self-consistent iterations was assumed when the total energy difference between cycles was $<1.0 \times 10^{-7}$ eV. The basis calculations were performed in the unit cell of $Ca_3Ti_2O_7$ with 48 atoms. Monkhorst–Pack $k$-point meshes[39] (9×9×3) with a Gaussian searing of 0.05 eV were used to sample the Brillouin zone.

**References**
1. Nataf, G. F. *et al.* Domain-wall engineering and topological defects in ferroelectric and ferroelastic materials. *Nat. Rev. Phys.* **2**, 634–648 (2020).
2. Catalan, G., Seidel, J., Ramesh, R. & Scott, J. F. Domain wall nanoelectronics. *Rev. Mod. Phys.* **84**, 119 (2012).
3. Vasudevan, R. K. *et al.* Domain wall conduction and polarization-mediated transport in ferroelectrics. *Adv. Funct. Mater.* **23**, 2592–2616 (2013).
4. Wei, X.-K. *et al.* Ferroelectric translational antiphase boundaries in nonpolar materials. *Nat. Commun.* **5**, 1–8 (2014).
5. Yokota, H. *et al.* Direct evidence of polar nature of ferroelastic twin boundaries in $CaTiO_3$ obtained by second harmonic generation microscope. *Phys. Rev. B* **89**, 144109 (2014).
6. Seidel, J. *et al.* Conduction at domain walls in oxide multiferroics. *Nat. Mater.* **8**, 229–234 (2009).
7. Sharma, P. *et al.* Nonvolatile ferroelectric domain wall memory. *Sci. Adv.* **3**, e1700512 (2017).
8. Sluka, T., Tagantsev, A. K., Bednyakov, P. & Setter, N. Free-electron gas at charged domain walls in insulating $BaTiO_3$. *Nat. Commun.* **4**, 1–6 (2013).
9. Bednyakov, P. S., Sturman, B. I., Sluka, T., Tagantsev, A. K. & Yudin, P. V. Physics and applications of charged domain walls. *npj Comput. Mater.* **4**, 65 (2018).
10. Toledano, P. & Toledano, J. *Landau Theory Of Phase Transitions, The: Application To Structural, Incommensurate, Magnetic And Liquid Crystal Systems*. vol. 3 (World Scientific Publishing Company, 1987).
11. Choi, T. *et al.* Insulating interlocked ferroelectric and structural antiphase domain walls in multiferroic $YMnO_3$. *Nat. Mater.* **9**, 253 (2010).
12. Wu, W., Horibe, Y., Lee, N., Cheong, S.-W. & Guest, J. R. Conduction of topologically protected charged ferroelectric domain walls. *Phys. Rev. Lett.* **108**, 77203 (2012).

13. Meier, D. *et al.* Anisotropic conductance at improper ferroelectric domain walls. *Nat. Mater.* **11**, 284 (2012).
14. Mundy, J. A. *et al.* Functional electronic inversion layers at ferroelectric domain walls. *Nat. Mater.* **16**, 622 (2017).
15. Benedek, N. A. & Fennie, C. J. Hybrid improper ferroelectricity: a mechanism for controllable polarization-magnetization coupling. *Phys. Rev. Lett.* **106**, 107204 (2011).
16. Oh, Y. S., Luo, X., Huang, F.-T., Wang, Y. & Cheong, S.-W. Experimental demonstration of hybrid improper ferroelectricity and the presence of abundant charged walls in $(Ca, Sr)_3Ti_2O_7$ crystals. *Nat. Mater.* **14**, 407 (2015).
17. Kumagai, Y. & Spaldin, N. A. Structural domain walls in polar hexagonal manganites. *Nat. Commun.* **4**, 1–8 (2013).
18. Jungk, T., Hoffmann, Á., Fiebig, M. & Soergel, E. Electrostatic topology of ferroelectric domains in $YMnO_3$. *Appl. Phys. Lett.* **97**, 12904 (2010).
19. Chae, S. C. *et al.* Evolution of the domain topology in a ferroelectric. *Phys. Rev. Lett.* **110**, 167601 (2013).
20. Zhang, Q.-H. *et al.* Topology breaking of the vortex in multiferroic $Y_{0.67}Lu_{0.33}MnO_3$. *Appl. Phys. Lett.* **105**, 12902 (2014).
21. Zurbuchen, M. A. *et al.* Morphology, structure, and nucleation of out-of-phase boundaries (OPBs) in epitaxial films of layered oxides. *J. Mater. Res.* **22**, 1439–1471 (2007).
22. Zurbuchen, M. A. *et al.* Bismuth volatility effects on the perfection of $SrBi_2Nb_2O_9$ and $SrBi_2Ta_2O_9$ films. *Appl. Phys. Lett.* **82**, 4711–4713 (2003).
23. Pan, X. Q., Jiang, J. C., Theis, C. D. & Schlom, D. G. Domain structure of epitaxial $Bi_4Ti_3O_{12}$ thin films grown on (001) $SrTiO_3$ substrates. *Appl. Phys. Lett.* **83**, 2315–2317 (2003).
24. Tanaka, M. & Honjo, G. Electron optical studies of barium titanate single crystal films. *J. Phys. Soc. Japan* **19**, 954–970 (1964).
25. Fujimoto, F. Dynamical theory of electron diffraction in Laue-case, I. General theory. *J. Phys. Soc. Japan* **14**, 1558–1568 (1959).
26. Huang, F.-T. *et al.* Domain topology and domain switching kinetics in a hybrid improper ferroelectric. *Nat. Commun.* **7**, 11602 (2016).
27. Lee, M. H. *et al.* Hidden antipolar order parameter and entangled Néel-type charged domain walls in hybrid improper ferroelectrics. *Phys. Rev. Lett.* **119**, 157601 (2017).
28. Mitsui, T. & Furuichi, J. Domain Structure of Rochelle Salt and $KH_2PO_4$. *Phys. Rev.* **90**, 193 (1953).
29. Benedek, N. A., Mulder, A. T. & Fennie, C. J. Polar octahedral rotations: a path to new multifunctional materials. *J. Solid State Chem.* **195**, 11–20 (2012).
30. Amelinckx, S. & Van Landuyt, J. Transmission electron microscopy. (2003).
31. De Graef, M. *Introduction to conventional transmission electron microscopy*. (Cambridge University Press, 2003).
32. Huang, F.-T. *et al.* Topological defects at octahedral tilting plethora in bi-layered perovskites. *npj Quantum Mater.* **1**, 16017 (2016).
33. Rojac, T. *et al.* Domain-wall conduction in ferroelectric $BiFeO_3$ controlled by accumulation of charged defects. *Nat. Mater.* **16**, 322–327 (2017).

34. Ishizuka, K., Eilers, P. H. C. & Kogure, T. Optimal noise filters in high-resolution electron microscopy. *Micros. Today* **15**, 16–21 (2007).
35. Kresse, G. & Furthmüller, J. Efficient iterative schemes for ab initio total-energy calculations using a plane-wave basis set. *Phys. Rev. B* **54**, 11169 (1996).
36. Blöchl, P. E. Projector augmented-wave method. *Phys. Rev. B* **50**, 17953 (1994).
37. Perdew, J. P., Burke, K. & Ernzerhof, M. Generalized gradient approximation made simple. *Phys. Rev. Lett.* **77**, 3865 (1996).
38. Li, C. F. *et al.* Structural transitions in hybrid improper ferroelectric $Ca_3Ti_2O_7$ tuned by site-selective isovalent substitutions: A first-principles study. *Phys. Rev. B* **97**, 184105 (2018).
39. Monkhorst, H. J. & Pack, J. D. Special points for Brillouin-zone integrations. *Phys. Rev. B* **13**, 5188 (1976).
40. Li, G. J., Liu, X. Q., Lu, J. J., Zhu, H. Y. & Chen, X. M. Crystal structural evolution and hybrid improper ferroelectricity in Ruddlesden-Popper $Ca_{3-x}Sr_xTi_2O_7$ ceramics. *J. Appl. Phys.* **123**, (2018).
41. Momma, K. & Izumi, F. VESTA 3 for three-dimensional visualization of crystal, volumetric and morphology data. *J. Appl. Crystallogr.* **44**, 1272–1276 (2011).

**Figures**

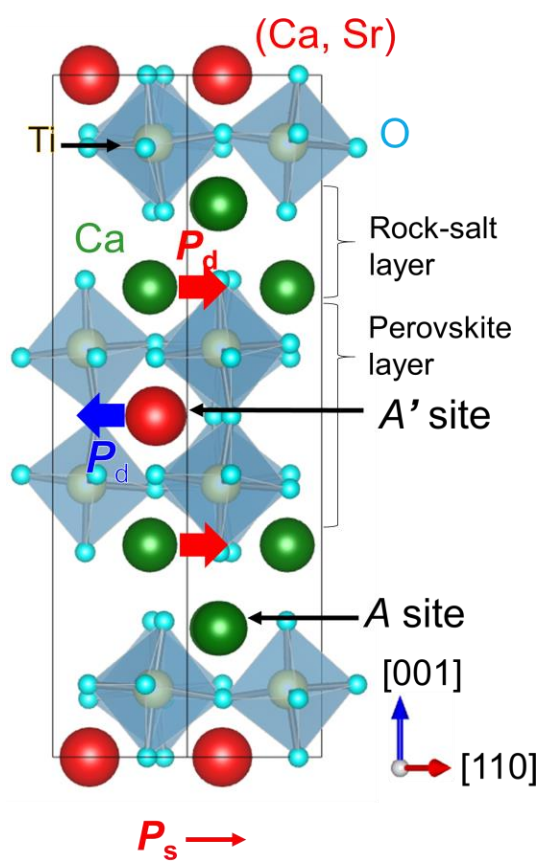

**Figure 1. Crystal structure viewed along the [1$\bar{1}$0] axis.** The $P_d$ and $P_s$ arrows represent the displacements and the spontaneous polarization directions, respectively. The crystal structure is based on Refs. [40,41]. The green and red spheres indicate the sites of the rock-salt and perovskite layers. The sites labeled A (rock-salt layer) and A' (perovskite layer) are cation sites calculating the substitution energy using the density-functional theory, as explained in Supplementary Note 5.

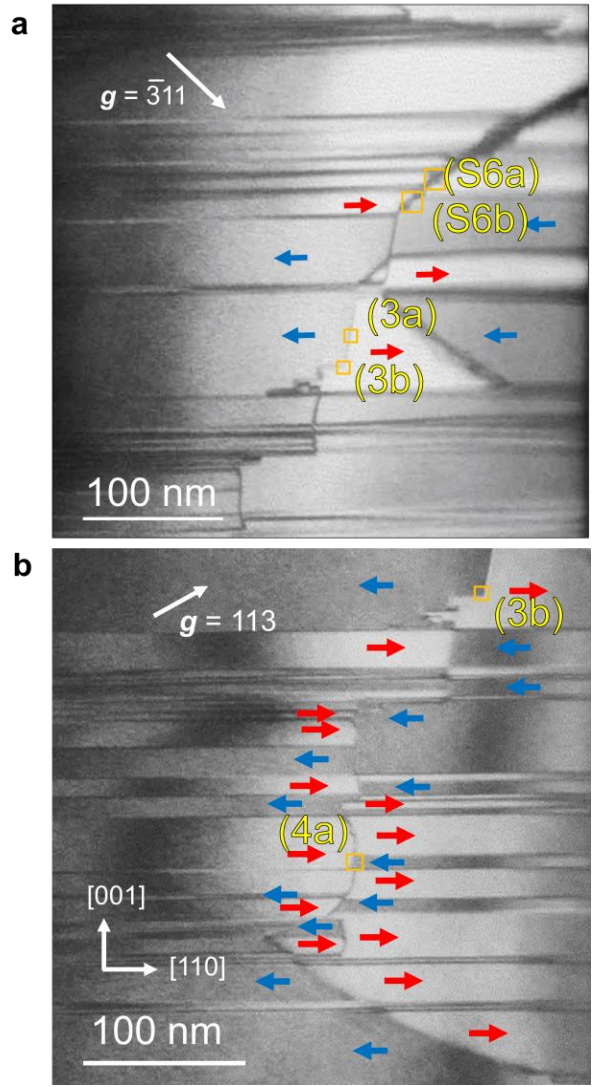

**Figure 2. Observation of charged structural defects using dark-field imaging. a** Dark-field image for $g = \bar{3}11$ in the $(1\bar{1}0)$ plane. **b** Dark-field image for $g = 113$. The areas indicated by squares show the locations at which the EDS images in Figures 3 and 4 were acquired. The red and blue arrows show the projected polarization directions in the ferroelectric domains. Notably, the area labeled 3b has the same position in panels **a** and **b**. The straight lines indicate out-of-phase boundaries with translational shifts, whereas the broad boundaries correspond to regions with SrTiO$_3$ structures, as explained in the main text.

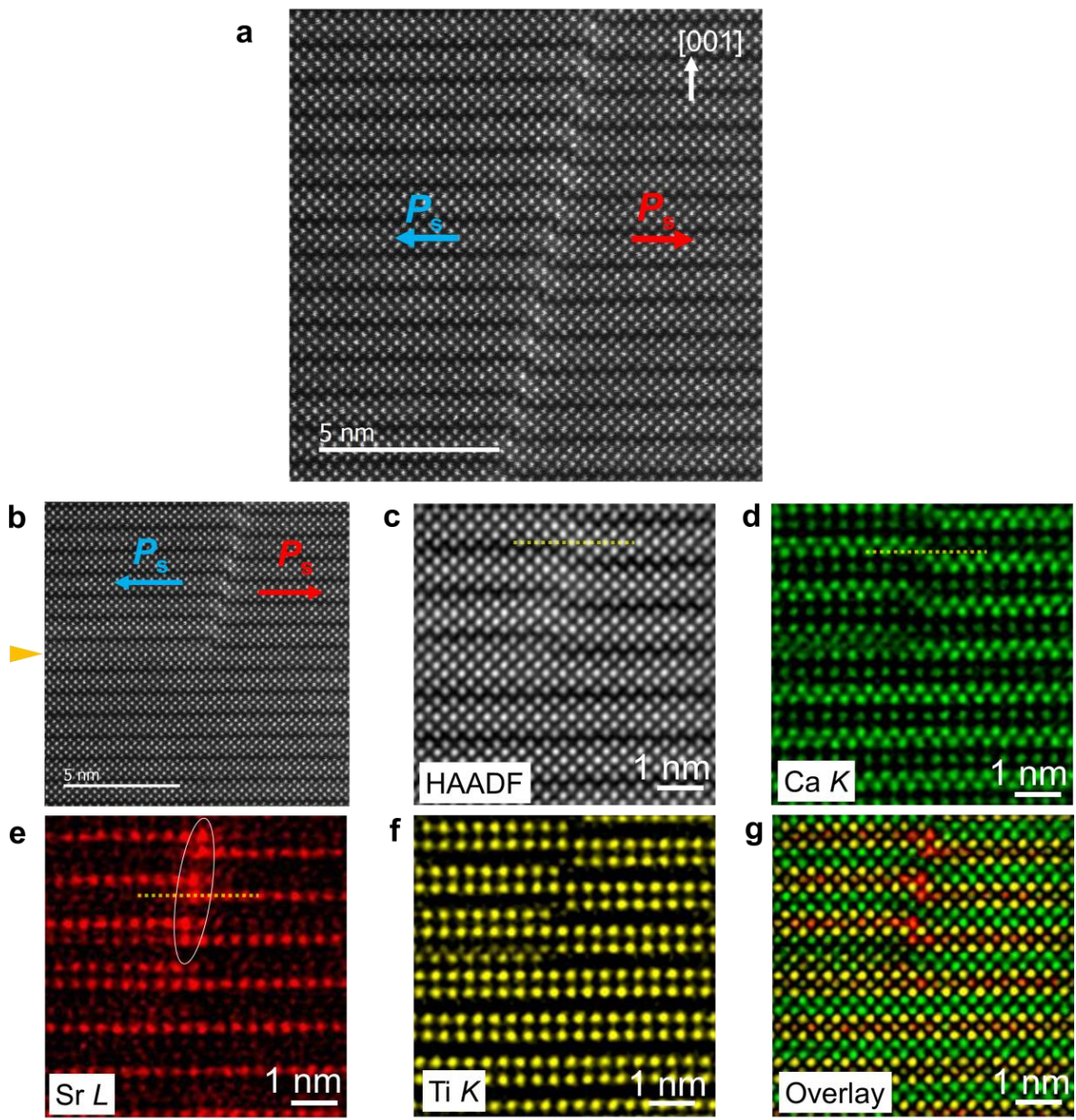

**Figure 3. Microstructure of a charged out-of-phase boundary. a** HAADF-STEM image of a charged domain boundary in the area indicated in Figure 2. The contrast level of the dark-field image determines the projected polarization direction $P_s$. **b** HAADF-STEM image of the edge of the charged domain boundary. **c**–**g** HAADF-STEM image and corresponding elemental maps of the edge of the boundary. The labels *L* and *K* represent the absorption edges for EDS mapping. The segregation of Sr atoms is shown in panel **e**.

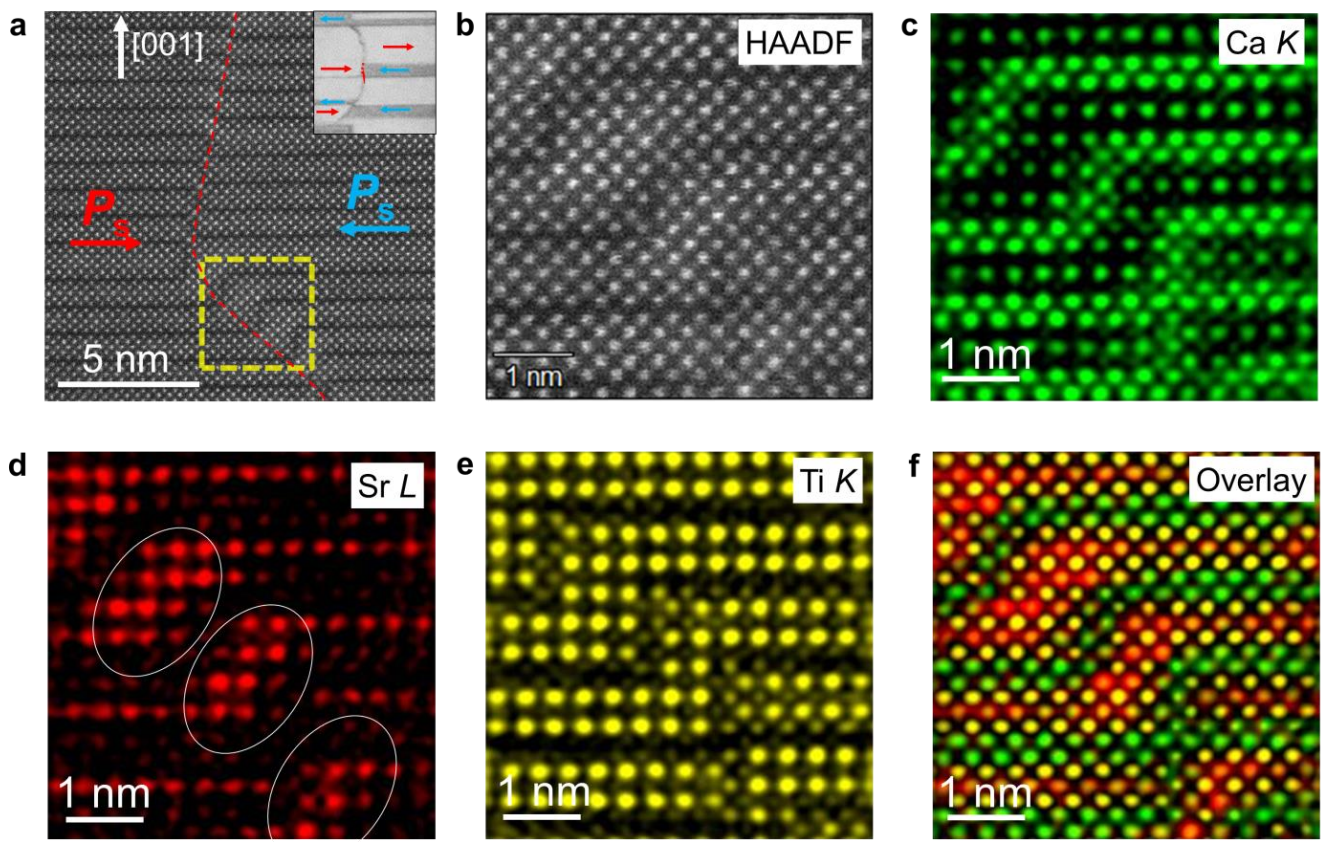

**Figure 4. Observation of rotated out-of-phase domains. a**, **b** HAADF-STEM images showing boundary structures. The square area indicated by a dashed yellow line is shown in high magnification in panel **b**. The inset shows the magnified dark-field image of Figure 2. The dashed line denotes a charged domain boundary. $P_s$ indicates the direction of the electric polarization in the domains. The corresponding elemental maps are shown in panels **c**–**f**.

15